\documentclass[aps,prd,twocolumn,superscriptaddress,amsmath,showpacs]{revtex4}

\usepackage{graphicx}% Include figure files
\usepackage{dcolumn}% Align table columns on decimal point
\usepackage{bm}% bold math
\usepackage{axodraw}
\usepackage{multirow} 
	
\newcommand{\Kstar} {\ensuremath{{K^*}^0}}
\newcommand{\KSpipi} {\ensuremath{K^0_S\to \pi^+ \pi^-}}
\newcommand{\KstarKpi} {\ensuremath{{K^*}^0\to K^+ \pi^-}}
\newcommand{\BsJpsiKstar} {\ensuremath{B^0_s \to J/\psi \,{K^*}^0}}
\newcommand{\BsJpsiKshort} {\ensuremath{B^0_s \to J/\psi \, K^0_S}}
\newcommand{\Jpsimumu} {\ensuremath{J/\psi \to \mu^+\mu^-}}

\begin{document}
\title{
{Observation of $B^0_s \to J/\psi\,{K^*(892)}^0$ and $B^0_s \to J/\psi \, K^0_S$ decays}}
\affiliation{Institute of Physics, Academia Sinica, Taipei, Taiwan 11529, Republic of China} 
\affiliation{Argonne National Laboratory, Argonne, Illinois 60439, USA} 
\affiliation{University of Athens, 157 71 Athens, Greece} 
\affiliation{Institut de Fisica d'Altes Energies, ICREA, Universitat Autonoma de Barcelona, E-08193, Bellaterra (Barcelona), Spain} 
\affiliation{Baylor University, Waco, Texas 76798, USA} 
\affiliation{Istituto Nazionale di Fisica Nucleare Bologna, $^z$University of Bologna, I-40127 Bologna, Italy} 
\affiliation{University of California, Davis, Davis, California 95616, USA} 
\affiliation{University of California, Los Angeles, Los Angeles, California 90024, USA} 
\affiliation{Instituto de Fisica de Cantabria, CSIC-University of Cantabria, 39005 Santander, Spain} 
\affiliation{Carnegie Mellon University, Pittsburgh, Pennsylvania 15213, USA} 
\affiliation{Enrico Fermi Institute, University of Chicago, Chicago, Illinois 60637, USA}
\affiliation{Comenius University, 842 48 Bratislava, Slovakia; Institute of Experimental Physics, 040 01 Kosice, Slovakia} 
\affiliation{Joint Institute for Nuclear Research, RU-141980 Dubna, Russia} 
\affiliation{Duke University, Durham, North Carolina 27708, USA} 
\affiliation{Fermi National Accelerator Laboratory, Batavia, Illinois 60510, USA} 
\affiliation{University of Florida, Gainesville, Florida 32611, USA} 
\affiliation{Laboratori Nazionali di Frascati, Istituto Nazionale di Fisica Nucleare, I-00044 Frascati, Italy} 
\affiliation{University of Geneva, CH-1211 Geneva 4, Switzerland} 
\affiliation{Glasgow University, Glasgow G12 8QQ, United Kingdom} 
\affiliation{Harvard University, Cambridge, Massachusetts 02138, USA} 
\affiliation{Division of High Energy Physics, Department of Physics, University of Helsinki and Helsinki Institute of Physics, FIN-00014, Helsinki, Finland} 
\affiliation{University of Illinois, Urbana, Illinois 61801, USA} 
\affiliation{The Johns Hopkins University, Baltimore, Maryland 21218, USA} 
\affiliation{Institut f\"{u}r Experimentelle Kernphysik, Karlsruhe Institute of Technology, D-76131 Karlsruhe, Germany} 
\affiliation{Center for High Energy Physics: Kyungpook National University, Daegu 702-701, Korea; Seoul National University, Seoul 151-742, Korea; Sungkyunkwan University, Suwon 440-746, Korea; Korea Institute of Science and Technology Information, Daejeon 305-806, Korea; Chonnam National University, Gwangju 500-757, Korea; Chonbuk National University, Jeonju 561-756, Korea} 
\affiliation{Ernest Orlando Lawrence Berkeley National Laboratory, Berkeley, California 94720, USA} 
\affiliation{University of Liverpool, Liverpool L69 7ZE, United Kingdom} 
\affiliation{University College London, London WC1E 6BT, United Kingdom} 
\affiliation{Centro de Investigaciones Energeticas Medioambientales y Tecnologicas, E-28040 Madrid, Spain} 
\affiliation{Massachusetts Institute of Technology, Cambridge, Massachusetts 02139, USA} 
\affiliation{Institute of Particle Physics: McGill University, Montr\'{e}al, Qu\'{e}bec, Canada H3A~2T8; Simon Fraser University, Burnaby, British Columbia, Canada V5A~1S6; University of Toronto, Toronto, Ontario, Canada M5S~1A7; and TRIUMF, Vancouver, British Columbia, Canada V6T~2A3} 
\affiliation{University of Michigan, Ann Arbor, Michigan 48109, USA} 
\affiliation{Michigan State University, East Lansing, Michigan 48824, USA}
\affiliation{Institution for Theoretical and Experimental Physics, ITEP, Moscow 117259, Russia}
\affiliation{University of New Mexico, Albuquerque, New Mexico 87131, USA} 
\affiliation{Northwestern University, Evanston, Illinois 60208, USA} 
\affiliation{The Ohio State University, Columbus, Ohio 43210, USA} 
\affiliation{Okayama University, Okayama 700-8530, Japan} 
\affiliation{Osaka City University, Osaka 588, Japan} 
\affiliation{University of Oxford, Oxford OX1 3RH, United Kingdom} 
\affiliation{Istituto Nazionale di Fisica Nucleare, Sezione di Padova-Trento, $^{aa}$University of Padova, I-35131 Padova, Italy} 
\affiliation{LPNHE, Universite Pierre et Marie Curie/IN2P3-CNRS, UMR7585, Paris, F-75252 France} 
\affiliation{University of Pennsylvania, Philadelphia, Pennsylvania 19104, USA}
\affiliation{Istituto Nazionale di Fisica Nucleare Pisa, $^{bb}$University of Pisa, $^{cc}$University of Siena and $^{dd}$Scuola Normale Superiore, I-56127 Pisa, Italy} 
\affiliation{University of Pittsburgh, Pittsburgh, Pennsylvania 15260, USA} 
\affiliation{Purdue University, West Lafayette, Indiana 47907, USA} 
\affiliation{University of Rochester, Rochester, New York 14627, USA} 
\affiliation{The Rockefeller University, New York, New York 10065, USA} 
\affiliation{Istituto Nazionale di Fisica Nucleare, Sezione di Roma 1, $^{ee}$Sapienza Universit\`{a} di Roma, I-00185 Roma, Italy} 

\affiliation{Rutgers University, Piscataway, New Jersey 08855, USA} 
\affiliation{Texas A\&M University, College Station, Texas 77843, USA} 
\affiliation{Istituto Nazionale di Fisica Nucleare Trieste/Udine, I-34100 Trieste, $^{ff}$University of Trieste/Udine, I-33100 Udine, Italy} 
\affiliation{University of Tsukuba, Tsukuba, Ibaraki 305, Japan} 
\affiliation{Tufts University, Medford, Massachusetts 02155, USA} 
\affiliation{University of Virginia, Charlottesville, VA  22906, USA}
\affiliation{Waseda University, Tokyo 169, Japan} 
\affiliation{Wayne State University, Detroit, Michigan 48201, USA} 
\affiliation{University of Wisconsin, Madison, Wisconsin 53706, USA} 
\affiliation{Yale University, New Haven, Connecticut 06520, USA} 
\author{T.~Aaltonen}
\affiliation{Division of High Energy Physics, Department of Physics, University of Helsinki and Helsinki Institute of Physics, FIN-00014, Helsinki, Finland}
\author{B.~\'{A}lvarez~Gonz\'{a}lez$^v$}
\affiliation{Instituto de Fisica de Cantabria, CSIC-University of Cantabria, 39005 Santander, Spain}
\author{S.~Amerio}
\affiliation{Istituto Nazionale di Fisica Nucleare, Sezione di Padova-Trento, $^{aa}$University of Padova, I-35131 Padova, Italy} 

\author{D.~Amidei}
\affiliation{University of Michigan, Ann Arbor, Michigan 48109, USA}
\author{A.~Anastassov}
\affiliation{Northwestern University, Evanston, Illinois 60208, USA}
\author{A.~Annovi}
\affiliation{Laboratori Nazionali di Frascati, Istituto Nazionale di Fisica Nucleare, I-00044 Frascati, Italy}
\author{J.~Antos}
\affiliation{Comenius University, 842 48 Bratislava, Slovakia; Institute of Experimental Physics, 040 01 Kosice, Slovakia}
\author{G.~Apollinari}
\affiliation{Fermi National Accelerator Laboratory, Batavia, Illinois 60510, USA}
\author{J.A.~Appel}
\affiliation{Fermi National Accelerator Laboratory, Batavia, Illinois 60510, USA}
\author{A.~Apresyan}
\affiliation{Purdue University, West Lafayette, Indiana 47907, USA}
% ...
\collaboration{CDF Collaboration\footnote{With visitors from $^a$University of Massachusetts Amherst, Amherst, Massachusetts 01003,
$^b$Istituto Nazionale di Fisica Nucleare, Sezione di Cagliari, 09042 Monserrato (Cagliari), Italy,
$^c$University of California Irvine, Irvine, CA  92697, 
$^d$University of California Santa Barbara, Santa Barbara, CA 93106
$^e$University of California Santa Cruz, Santa Cruz, CA  95064,
$^f$CERN,CH-1211 Geneva, Switzerland,
$^g$Cornell University, Ithaca, NY  14853, 
$^h$University of Cyprus, Nicosia CY-1678, Cyprus, 
$^i$University College Dublin, Dublin 4, Ireland,
$^j$University of Fukui, Fukui City, Fukui Prefecture, Japan 910-0017,
$^k$Universidad Iberoamericana, Mexico D.F., Mexico,
$^l$Iowa State University, Ames, IA  50011,
$^m$University of Iowa, Iowa City, IA  52242,
$^n$Kinki University, Higashi-Osaka City, Japan 577-8502,
$^o$Kansas State University, Manhattan, KS 66506,
$^p$University of Manchester, Manchester M13 9PL, England,
$^q$Queen Mary, University of London, London, E1 4NS, England,
$^r$Muons, Inc., Batavia, IL 60510,
$^s$Nagasaki Institute of Applied Science, Nagasaki, Japan, 
$^t$National Research Nuclear University, Moscow, Russia,
$^u$University of Notre Dame, Notre Dame, IN 46556,
$^v$Universidad de Oviedo, E-33007 Oviedo, Spain, 
$^w$Texas Tech University, Lubbock, TX  79609, 
$^x$Universidad Tecnica Federico Santa Maria, 110v Valparaiso, Chile,
$^y$Yarmouk University, Irbid 211-63, Jordan,
$^{gg}$On leave from J.~Stefan Institute, Ljubljana, Slovenia, 
}}
\noaffiliation

%\author{The CDF Collaboration
%  \it http://www-cdf.fnal.gov \rm }
%\noaffiliation

%\date{\today}

\begin{abstract}
 We report  the first observation of two Cabibbo-suppressed decay modes of the $B^0_s$ meson.  
 Using a  sample of  $p\bar{p}$ collisions at $\sqrt{s}=1.96\, \rm TeV$
  corresponding to $5.9\, \rm fb^{-1}$ of integrated luminosity collected with the CDF\,II detector at the Fermilab Tevatron,
 we search for new $B^0_s$ decay modes in a sample of events containing $J/\psi~\to~\mu^+\mu^-$ decays.  We reconstruct  a $B^0_s \to J/\psi \,{K^*(892)}^0$ signal with ${K^*(892)}^0\to K^+\pi^-$, observing a yield of $151\pm 25$ events with a statistical significance of $8.0\sigma$.  We also reconstruct a $B^0_s \to J/\psi \,K^0_S$ signal  with $K^0_S\to \pi^+\pi^-$, observing a yield of $64\pm 14$ events with a statistical significance of $7.2\sigma$.  From these yields, we extract the branching ratios ${\mathcal B}(B^0_s \rightarrow J/\psi \,{K^*(892)}^0)= (8.3 \pm 3.8)\times10^{-5}$ 
and ${\mathcal B}(B^0_s \rightarrow J/\psi\, K^0)= (3.5\pm 0.8)\times10^{-5}$, where statistical, systematic, and fragmentation-fraction uncertainties are included in the combined uncertainty.

\end{abstract}

%% add PACS numbers


\maketitle

%\linenumbers

%\newpage
\setcounter{secnumdepth}{1}

\section{Introduction}

This paper presents the first observation
 of the Cabibbo-suppressed decays $B^0_s\rightarrow J/\psi {K^*}^0(892)$ and \BsJpsiKshort\ (and the corresponding charge conjugate decays) using a sample derived from an integrated luminosity of $5.9\, {\rm fb}^{-1}$ of proton-antiproton collisions at $\sqrt{s}=1.96 \, \rm TeV$ produced at the Fermilab Tevatron. In addition to isolating these signals, we normalize the observed yields to the corresponding Cabibbo-favored $B^0$ decay modes ($B^0\to J/\psi \,{K^*}^0$, where ${K^*}^0$ refers to ${K^*}^0(892)$, and $B^0\to J/\psi \,K^0_S$) to extract the branching ratios for these newly observed $B^0_s$ decay modes using the relation

\begin{equation} \label{eq:BR}
\frac{{\mathcal B}(B^0_s \rightarrow J/\psi K)}{{\mathcal B}(B^0 \rightarrow J/\psi K)}=A_{rel}\frac{f_d}{f_s}\frac{N(B^0_s\rightarrow J/\psi K)}{N(B^0\rightarrow J/\psi K)},   
\end{equation}

\noindent
where $K$ represents $K^0_S$ or ${K^*}^0$, $A_{rel}$ is the relative acceptance, $f_s/f_d$ is the ratio of fragmentation fractions and $N(B^0_s \rightarrow J/\psi K)/N(B^0 \rightarrow J/\psi K)$ is the measured ratio of yields.

In the na\"{i}ve spectator model, the ratio of branching ratios is given by the ratio of the squares of the Cabibbo-Kobayashi-Maskawa (CKM) elements 

\begin{equation}
\frac{{\mathcal B}(B^0_s \rightarrow J/\psi K)}{{\mathcal B}(B^0 \rightarrow J/\psi K)}= \frac{|V_{cd}|^2}{|V_{cs}|^2} = 0.051 \pm 0.006, 
\label{eqratio}
\end{equation}
which is derived from $|V_{cd}| = 0.230\pm 0.011$ and $|V_{cs}| = 1.023\pm 0.036$ \cite{pdg}.   
Assuming a relative acceptance $A_{rel}$ of unity, we estimate a ratio of yields. 
The value for $f_s/f_d$ is extracted from 
the 
most recent CDF measurement~\cite{fracCDF} of $f_s/(f_u+f_d)\times {\mathcal B}(D_s \rightarrow \phi \pi )$ and $f_u/f_d$ 
along with the current world-average value~\cite{pdg} for ${\mathcal B}(D_s \rightarrow \phi \pi )$. 
Combining the value $f_s/f_d$ = 0.269 $\pm$ 0.033 with Eq.~\ref{eqratio} yields

\begin{equation}
\begin{split}
\frac{N(B^0_s\rightarrow J/\psi K)}{N(B^0\rightarrow J/\psi K)}&= \frac{{\mathcal B}(B^0_s \rightarrow J/\psi K)}{{\mathcal B}(B^0 \rightarrow J/\psi K)}\frac{f_s}{f_d}\frac{1}{A_{rel}} \\
&= 0.014 \pm 0.002.
\end{split}
\end{equation}

While the result holds only in the simple spectator case, it provides useful guidance that we might expect one to two 
Cabibbo-suppressed $B^0_s\rightarrow J/\psi K$ events for every 100 Cabibbo-favored $B^0 \rightarrow J/\psi K$ events.

With the establishment of the decay modes presented here, future measurements can be considered that will further aid 
our experimental investigation into the physics of the $B^0_s$ system.  The success of the CKM three-generation description of charge conjugation-parity inversion (\textit {CP}) violation \cite{ckm} in the bottom and kaon sectors has continued to motivate additional, more precise tests of \textit{CP} violation in the flavor sector.  In recent years, attention has turned to the $B^0_s$ meson as new territory to explore 
the possibility of non-standard-model contributions, specifically in the CKM matrix element $V_{ts}$. 
Precise measurement of the frequency of $B^0_s$ flavor oscillations \cite{mixing} has significantly limited the magnitude of 
new physics amplitudes. However, possible large new physics
phases remain poorly constrained.   

Cabibbo-suppressed $B^0_s$ modes could provide complementary information on the $B^0_s$ mixing phase and on the width difference $\Delta \Gamma_{B^0_s} = \Gamma_{B^0_{sL}} - \Gamma_{B^0_{sH}}$ where $\Gamma_{B^0_{sL}} (\Gamma_{B^0_{sH}})$ is the width of the light, even (heavy, odd) $B^0_s$ \textit{CP} eigenstate~\cite{pdg}.   The decay $B^0_s \to J/\psi \,{K^*(892)}^0$ is a pseudoscalar to vector-vector transition and can be used to help disentangle penguin contributions in $B^0_s \to J/\psi \,\phi$ \cite{fleischer}. With a sufficiently large data sample, it would be possible to measure $\Delta \Gamma_{B^0_s}$ and the polarization amplitudes. Furthermore, the Cabibbo-suppressed decay $B^0_s \to J/\psi \,K^0_S$ is a \textit {CP}-odd final state (ignoring \textit{CP} violation in the kaon system) and therefore a measurement of the lifetime in this decay mode is a direct measure of $\Gamma_{B^0_{sH}} = 1/\tau_{B^0_{sH}}$.  With a larger data sample, a tagged \textit{CP} asymmetry analysis of the \BsJpsiKshort \ mode, 
in conjunction with our precise knowledge of \textit{CP} violation in $B^0 \to J/\psi \,K^0_S$, can yield information
 on the angle $\gamma$ of the unitarity triangle \cite{gamma}.

After a description of the detector, data sample, and simulated samples utilized here, we describe the $B^0_s \rightarrow J/\psi {K^*}^0(892)$ analysis in Sec. III, followed by the $B^0_s \rightarrow J/\psi K^0_S$ analysis in Sec. IV. Section V then describes the acceptance calculation for both modes, followed by the results in Sec. VI. 

\section{CDF detector,  Data, and Monte Carlo samples}
The data used in these analyses correspond to an integrated luminosity of 5.9\,fb$^{-1}$ and were collected by the CDF II detector from March 2002 to February 2010 using di-muon triggers.  The CDF II detector is a general purpose, 
cylindrically symmetric detector. A more detailed description can be found elsewhere~\cite{cdf}. The sub-detectors relevant for these analyses are briefly discussed here. Charged particle trajectories (tracks) are measured by a system comprised of eight layers of silicon microstrip detector (SVX) and an open-cell wire drift chamber (COT), both immersed in a 1.4\,T axial magnetic field. The silicon detector~\cite{silis} extends from a radius of 1.5\,cm to 22\,cm and has a single-hit resolution of approximately 15\,$\mu$m. The COT drift chamber~\cite{COT} provides up to 96 measurements from radii of 40\,cm to 137\,cm and covers 
the range $|\eta| \leq$1~\cite{eta}.  Combined COT+SVX charged particle momentum resolution is 
$\sigma_{p_T}/{{p_T}^2} = 0.07\%$\,[GeV/$c\,]^{-1}$.  Outside the calorimeters reside  four layers of planar drift chambers~\cite{Muons1} (CMU) that detect muons with transverse momentum $p_T >$1.4\,GeV/$c$ within $|\eta|<0.6$. Additional chambers and scintillators~\cite{Muons2} (CMX) cover 0.6$<|\eta|<$1.0 for muons with $p_T >$2\,GeV/$c$.

The di-muon triggers collect a sample of \Jpsimumu \ candidates. At the first level of a three-level trigger system, an electronic track processor (XFT)~\cite{XFT} uses COT information to find tracks and extrapolate~\cite{XTRP} those with $p_T >$1.5(2.0)\,GeV/$c$ to track segments in the CMU (CMX) muon-chambers. Events pass this first trigger level if two or more XFT tracks are matched to muon-chamber track segments. The second trigger level requires those tracks to have opposite charge and an appropriate opening angle in the plane transverse to the beamline. Finally, at level 3, full tracking information is used to reconstruct \Jpsimumu \ candidates. Events with a candidate in the mass range 2.7 to 4.0\,GeV/$c^2$ are accepted. 

To identify $B^0$ and $B^0_s$ decay candidates, we pair $J/\psi$ candidates with \KSpipi \ and \KstarKpi candidates. The reconstruction of \KSpipi\  and \KstarKpi\ candidates starts from pairs of oppositely-charged tracks fit to a common interaction point (vertex). In the \BsJpsiKshort\ analysis, we reconstruct two tracks as pions and combine them to define a $K^0_S$ candidate, where the invariant mass of the two pions is constrained to the known $K^0_S$ mass~\cite{pdg}. In the \BsJpsiKstar\ analysis, we reconstruct the \Kstar\ candidate from the combination of a $\pi$ and a $K$. If two \Kstar\  candidates are reconstructed with the same tracks, with the only difference that the kaon and pion hypotheses are interchanged, we select the \Kstar \ candidate whose mass is closest to the pole value of 896\,MeV/$c^2$. We perform a kinematic fit of each $B$ candidate where the final state tracks  are constrained to come from a common decay point and the invariant mass of the muon pair is constrained to the known $J/\psi$ mass~\cite{pdg}.   These preliminary selection criteria for $B^0$ and $B_s^0$ candidates are listed in Table~\ref{tabCutsKS}.   Additional selection criteria optimized for the individual channels are described in Secs. III and IV.

Simulated samples of $B^0$ and $B^0_s$ decays are used to optimize event selection, 
model signal distributions, and assess systematic uncertainties.   
For our default Monte Carlo simulation (MC) samples, we generate  single $b$ hadrons according to the 
predicted next-to-leading order QCD calculation \cite{nde}.  For systematic studies, we also 
generate single $b$ hadrons according to  
momentum and rapidity spectra measured by CDF~\cite{cdf}.
These hadrons are then decayed using the \sc evtgen \rm
package \cite{evtgen} and then fed into a \sc geant \rm simulation of the CDF 
detector \cite{geant}.  The simulated data are then processed and reconstructed in the same
manner as the detector data. In the case of $J/\psi {K^*}^0$ mode, it 
is necessary to specify the polarization parameters in the simulation.  For both $B^0$ and 
$B^0_s$, we use transversity basis \cite{transversity}
polarization amplitudes $|A_0|^2 = 0.6$ and $|A_\perp|^2=0.22$, which are similar to the PDG values of
$|A_0|^2 = 0.571\pm 0.008$ and $|A_\perp|^2=0.22\pm 0.013$ \cite{pdg}. For systematic acceptance studies, MC samples 
with other polarization values were generated.

In all of the MC samples generated, and throughout the analyses presented below, we assume that there is no {\it CP} 
violation in $B^0_s$ mixing or decay. We additionally assume that equal numbers of $B^0$ and $\bar{B^0}$ mesons, 
as well as equal numbers of $B^0_s$ and $\bar{B^0_s}$ mesons, are produced in the $p\bar{p}$ collisions. 
From these assumptions, this untagged analysis is insensitive to {\it CP} violation $B^0$ decays, and the 
width difference in the $B^0_s$ system is given by $\Delta \Gamma_{B^0_s}$.

\section{$B^0_s \rightarrow J/\psi \,{K^{*}}^0$ analysis}
We optimize the selection criteria to provide the highest likelihood for evidence of this mode. This is done by maximizing $S/(1.5+\sqrt B)$, where $S$ refers to the number of signal events and $B$ is the number of background events in the signal region. Reference~\cite{punzi} demonstrates that this 
quantity is well suited for discovery. For the signal sample, a $B^0_s \rightarrow J/\psi \,{K^*}^0$ MC sample is used. 
For the background sample, we use  $J/\psi \, {K^*}^0$ candidate events from data with the requirement 
that the reconstructed candidate mass $M_B$ falls in the range
5.6\,GeV/$c^2< M_B <$ 5.8\,GeV/$c^2$.  
This ``upper sideband'' region contains events kinematically similar to the combinatorial background in the signal region and is not 
contaminated by residual signal events. We optimize simultaneously over the transverse momenta 
$p_T(\pi^-)$ and $p_T(K^+)$, the $B^0_s$ transverse decay length $L_{xy}(B^0_s)$, and the $B^0_s$ 
decay kinematic-fit probability. The final cuts we use are $p_T(\pi^-) > 1.5~{\rm GeV}/c$, $p_T (K^+) > 1.5~{\rm GeV}/c$, $L_{xy}(B^0_s) >$ 300 $\mu$m and fit probability greater than 10$^{-5}$.

Particle identification using specific ionization ($dE/dx$) in the COT was evaluated to further separate ${K^*}^0\rightarrow K^+\pi^-$ from  $\pi^+\pi^-$ and $K^+K^-$ backgrounds. Although further background reduction could be achieved, the corresponding reduction in  
signal efficiency rendered particle identification unprofitable, and we choose not to use it.  

We determine the $B^0_s$ and $B^0$ yields using a binned likelihood fit in the candidate masses. We model the signal contributions with templates composed of three Gaussians obtained from fits to $B^0$ MC. The two dominant, narrow Gaussians model detector resolution effects and also account for cases where the identities of the $\pi$ and $K$ from the \Kstar \ decay are interchanged. As mentioned above, we identify events where both $\pi$-$K$ and $K$-$\pi$ hypotheses pass the selection criteria and, in those cases, choose the combination closest to the $K^*(892)$ mass to ensure that candidates are not used twice. Approximately 10\% of $B\rightarrow J/\psi {K^*}^0$ events are reconstructed with the incorrect $\pi$-$K$ assignment. These events peak at the $B$ masses, but have a significantly broader width.

A wide Gaussian models misreconstructed signal events and other non-Gaussian resolution effects. 
The relative contributions, means, and widths of each Gaussian are fixed in the  fit. The $B^0_s$ templates used in the fit are identical to $B^0$ templates, except for a shift of 86.8 MeV/$c^2$ in the mean value of the three Gaussians. This value corresponds to the known~\cite{cdfmass, pdg} mass difference between $B^0_s$ and $B^0$. The MC slightly underestimates the 
mass resolution, so the widths of the two narrow Gaussians are multiplied by a scale factor common to the $B^0$ and $B^0_s$ templates, which is allowed to float in the fit. The scale factor is not applied to the third Gaussian since it is not expected to be affected by detector resolution effects as the other two are. Moreover, a common mass shift is added to the means of all Gaussian templates to account for a possible mass mismodeling in the MC. This mass shift is floating in the fit.

The $B^0_s \rightarrow J/\psi \,{K^*}^0$ analysis has three primary background contributions: events with random track combinations (combinatorics), 
partially reconstructed $b$ hadrons, and $B^0_s \rightarrow J/\psi \,\phi$ decays.  Combinatorial
background arises from sources such as a real $J/\psi$ plus two other tracks, where the $J/\psi$ could be either prompt or coming from a $B$ decay.  Another source arises from false $J/\psi$ candidates reconstructed from misidentified hadrons. The combinatorial background is modeled in the fit with an exponential function. 

Backgrounds from partially reconstructed $b$ hadrons come from multibody decays where a 
$\pi$, $K$, or $\gamma$ is not reconstructed, for example, the decay mode $B^0\rightarrow J/\psi {K^*}^0 \pi^0$.  We fit this background with two ARGUS functions~\cite{argus}, one for partially reconstructed $B^{0}$ and another for partially reconstructed $B^0_{s}$. The ARGUS function parameterization for $m$ $<$ $m_{0}$ is

\begin{equation}
f(m) = N_{1} \times \sqrt{ 1 - \frac{m^{2}}{m^{2}_{0}}} \times e^{-C m^{2}/m^{2}_{0}},
\end{equation}

\noindent
where $m_{0}$ is the mass cutoff, $C$ the decay constant, and $N_{1}$ is the normalization. The function is zero for $m$ $>$ $m_{0}$. The ARGUS function for partially reconstructed $B^{0}$ has a fixed mass cutoff of $m(B^{0}) - m(\pi^{0}) = 5.140$\,GeV/$c^{2}$ and the function for partially reconstructed $B^0_{s}$ has a fixed mass cutoff of $m(B^0_{s}) - m(\pi^{0}) = 5.220$\,GeV/$c^2$. The decay constants of the two functions are constrained to be the same, and the normalizations are independent. 
Each ARGUS function is convoluted with a Gaussian having a width of 12\,MeV/$c^2$ to account for detector resolution effects.

Since it is possible for $B^0_s \rightarrow J/\psi \, \phi$ candidates to pass the $J/\psi \, {K^*}^0$ reconstruction criteria, $B^0_s \rightarrow
J/\psi \, \phi$ must be considered as a background.  We use a template consisting of two Gaussians, extracted from simulation, to model
this background in the $J/\psi \, {K^*}^0$ fit, where both Gaussians are primarily modeling detector resolution effects.  We fix the widths, means, and relative contributions from each Gaussian in the final
fit. We multiply the constant width of the narrower Gaussian by the same scale factor used in the signal templates. We constrain the $B^0_s \rightarrow
J/\psi \, \phi$ contribution in the $J/\psi \, {K^*}^0$ fit by measuring the yield of $B^0_s \rightarrow J/\psi \, \phi$ in the data
using selection criteria efficient for reconstructing  $B^0_s \rightarrow J/\psi \, \phi$. We then use simulation to calculate the fraction of
those $J/\psi \, \phi$ events that would satisfy the $J/\psi \, {K^*}^0$ selection.

\begin{figure*}[!p]
  \includegraphics[angle=-90,width=8cm]{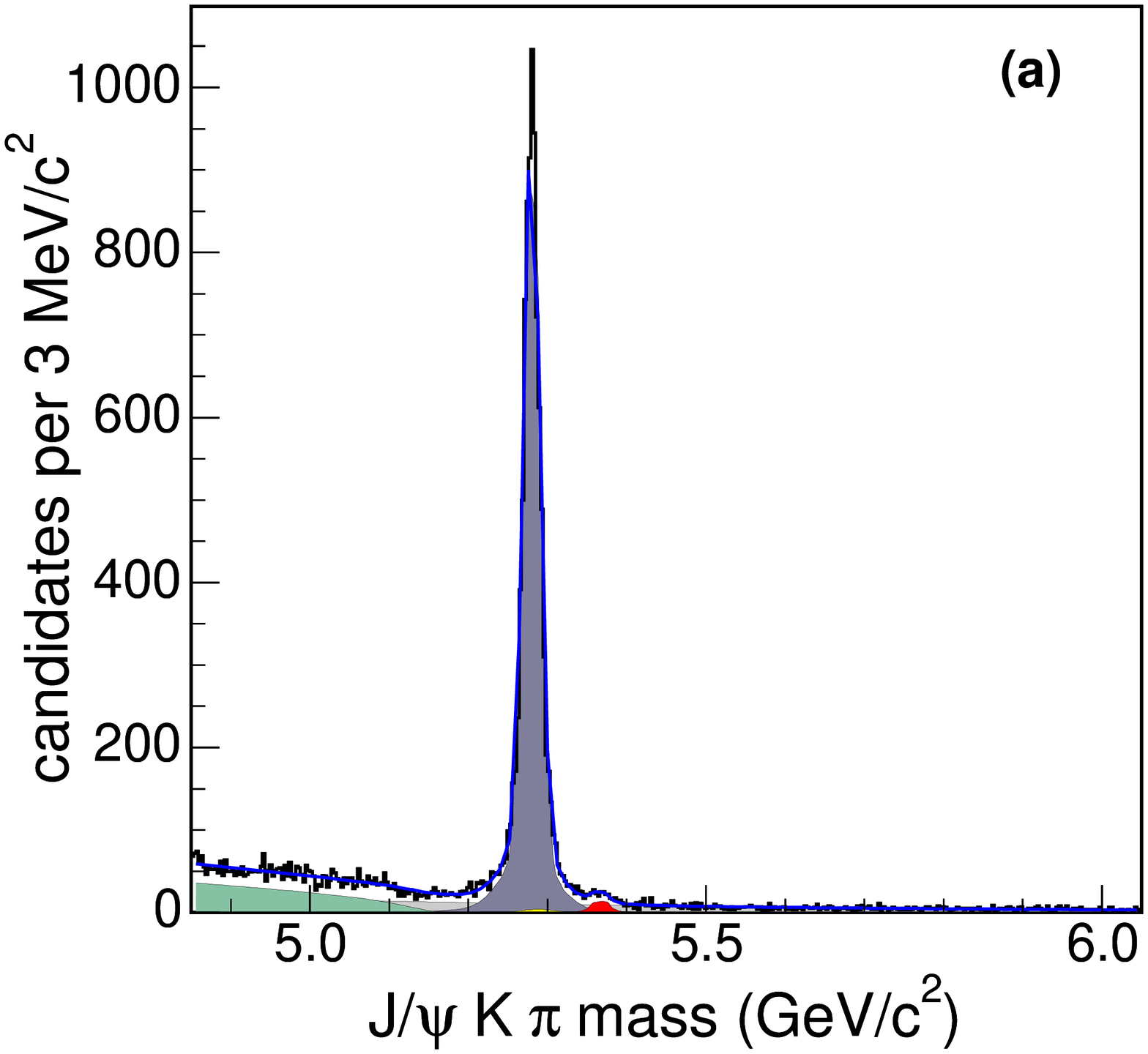}
  \includegraphics[angle=-90,width=8cm]{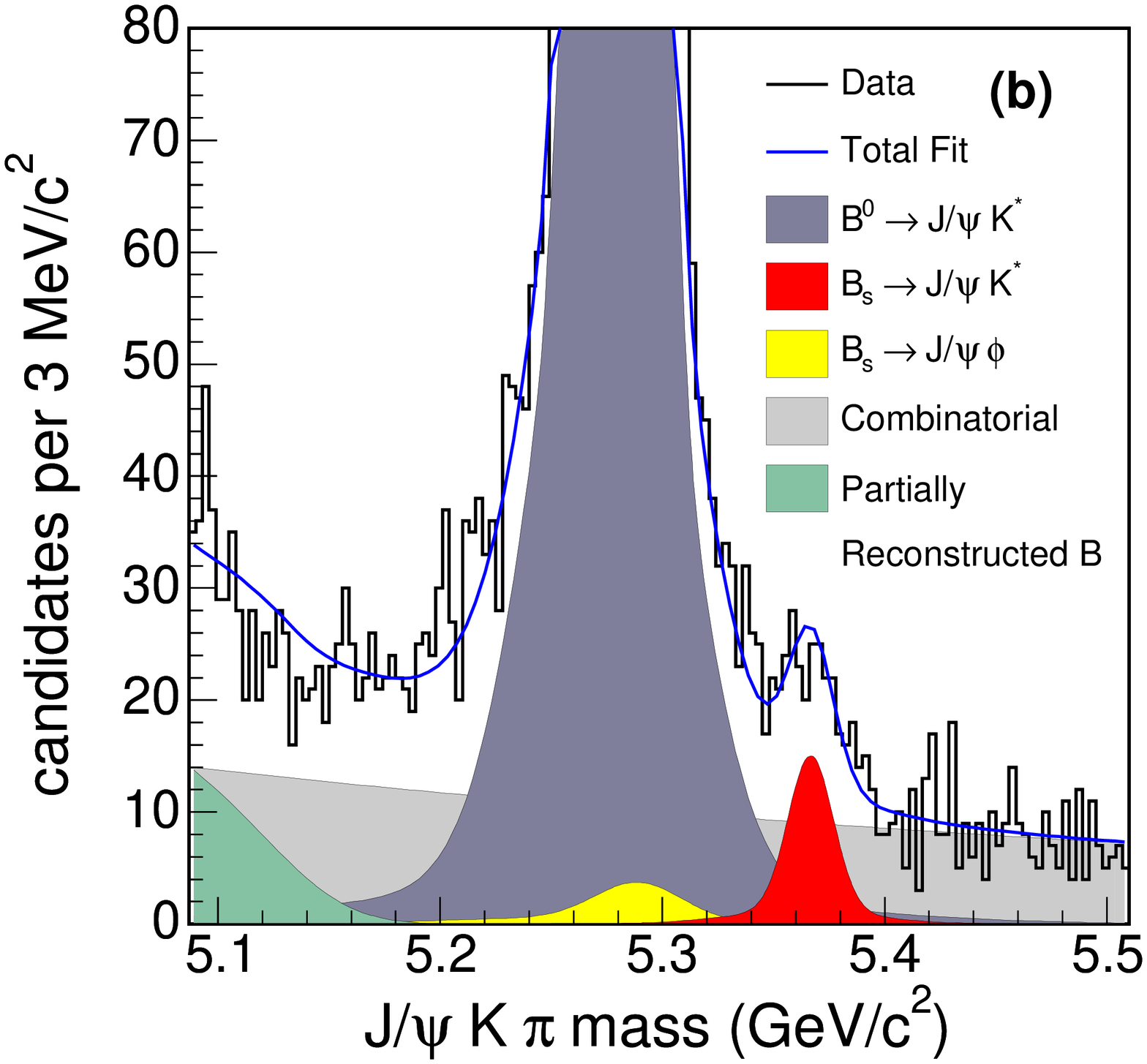}	
 \caption{\label{fig3}(a) Invariant mass distribution in data for $J/\psi \, {K^*}^0$ candidates and fit including the different contributions. 
(b) We enlarge the distribution in the signal region for more detail.}
\end{figure*}

We perform a binned log likelihood fit to the $J/\psi \, K \pi$ invariant mass distribution using the templates for signals and the background functions described above. The mass distributions in data for $J/\psi \, {K^*}^0$ candidates and the final fit appear in Fig.~\ref{fig3}. The yields for $B^0 \rightarrow J/\psi\, {K^*}^0$ and $B^0_s \rightarrow J/\psi \, {K^*}^0$ signal are 9530 $\pm$ 110 and 151 $\pm$ 25 respectively. The ratio $N(B^0_s \rightarrow J/\psi\, {K^*}^0)/N(B^0 \rightarrow J/\psi \, {K^*}^0)$ is 0.0159 $\pm$ 0.0022 (stat).

We determine the statistical significance of the $B^0_s \rightarrow J/\psi \, {K^*}^0$ signal 
 by fitting the mass distribution without the $B^0_s$ contribution (background-only hypothesis).  For likelihood ${\cal L}$, we interpret $-2\log{\cal L}$ as a $\chi^2$ distribution.  We use $\Delta \chi^2$ with one degree of freedom 
to determine that the probability of background fluctuations producing a comparable or greater signal is 8.9$\times$10$^{-16}$ or 8.0$\sigma$. This is the first observation of the $B^0_s \rightarrow J/\psi \, {K^*}^0$ decay.

We consider several sources of systematic uncertainty in the measured ratio of $N(B^0_s \rightarrow J/\psi \, {K^*}^0)/N(B^0 \rightarrow J/\psi \, {K^*}^0)$. The modeling of the $B^0$ and $B^0_s$ signal peaks can influence the measurement of the ratio. To quantify the effect of the mismodeling we repeat the fit using two Gaussian templates instead of three for the signal. The fit value of $N(B^0_s)/N(B^0)$ is shifted by 7$\times$10$^{-4}$.

We vary the input mass difference between $B^0$ and $B^0_s$ in the templates within its uncertainty of 0.7 MeV/$c^2$. The difference in 
$N(B^0_s \rightarrow J/\psi \, {K^*}^0)/N(B^0 \rightarrow J/\psi \, {K^*}^0)$ with the alternate templates is 2$\times$10$^{-5}$.  This is sufficiently small that we 
ascribe no systematic uncertainty for this effect.

The shape of the combinatorial background
is another source of systematic uncertainty. In this case, we use a power function instead of an exponential. We assign an additional systematic uncertainty of 2$\times$10$^{-4}$ to account for this effect.
 
In the likelihood fit, we allow the combinatorial background contribution to float. We performed a
study to evaluate how the ratio of yields depends upon the specific, arbitrary choice of the fit range. 
We compare the main fit, which allows the combinatorial background to float over the entire fit range, to a 
control case where the combinatorial contribution is fitted in the upper sideband and extrapolated to the full mass range prior to the final fit. 
Due to the difference in the result from these two methods, we include a systematic uncertainty of 0.0050 
on the $N(B^0_s \rightarrow J/\psi \, {K^*}^0)/N(B^0 \rightarrow J/\psi \, {K^*}^0)$ ratio.

To study the uncertainty in the $B^0_s \rightarrow J/\psi \, \phi$ contribution, we repeat the fit while doubling the 
fraction of $B^0_s \rightarrow J/\psi \, \phi$ candidates. The resulting shift of 2$\times$10$^{-4}$ is assigned as the
uncertainty in the $B^0_s \rightarrow J/\psi \, \phi$ contribution. 

We add the different systematic uncertainty contributions, summarized in Table~\ref{tabSysYields}, in quadrature resulting in a final value of $N(B^0_s \rightarrow J/\psi \, {K^*}^0)/N(B^0 \rightarrow J/\psi \, {K^*}^0)$ of 0.0159 $\pm$ 0.0022~(stat)~$\pm$~0.0050~(syst).

\section{$B^0_s \rightarrow J/\psi \, K^0_S$ analysis}
The $B^0_s \rightarrow J/\psi \, K^0_S$ decay has several differences compared to the $B^0_s \rightarrow J/\psi \, {K^*}^0$ decay. 
It contains a $K^0_S$, which has a  relatively long lifetime of $c\tau$=2.68 cm. We use the displacement between the reconstructed $K^0_S$ decay point and the reconstructed $B$ decay point in the event selection to reduce backgrounds such as $B^0_s \rightarrow J/\psi \, \phi$. Finally, as in the $B^0$ system, we expect the $B^0_s \rightarrow J/\psi \, K^0_S$ signal to be smaller than that of the  $B^0_s \rightarrow J/\psi \, {K^*}^0$ mode. Therefore, we use a Neural Network (NN) technique to take full advantage of all the kinematic variables and their correlations.  We use the NeuroBayes \cite{neurobayes} NN package.  The NN provides an output value close to +1 for signal-like events and near $-1$ for background-like events. 
 
\begin{figure*}[!p]
  \begin{center}
    \includegraphics[angle=-90,width=8cm,clip=]{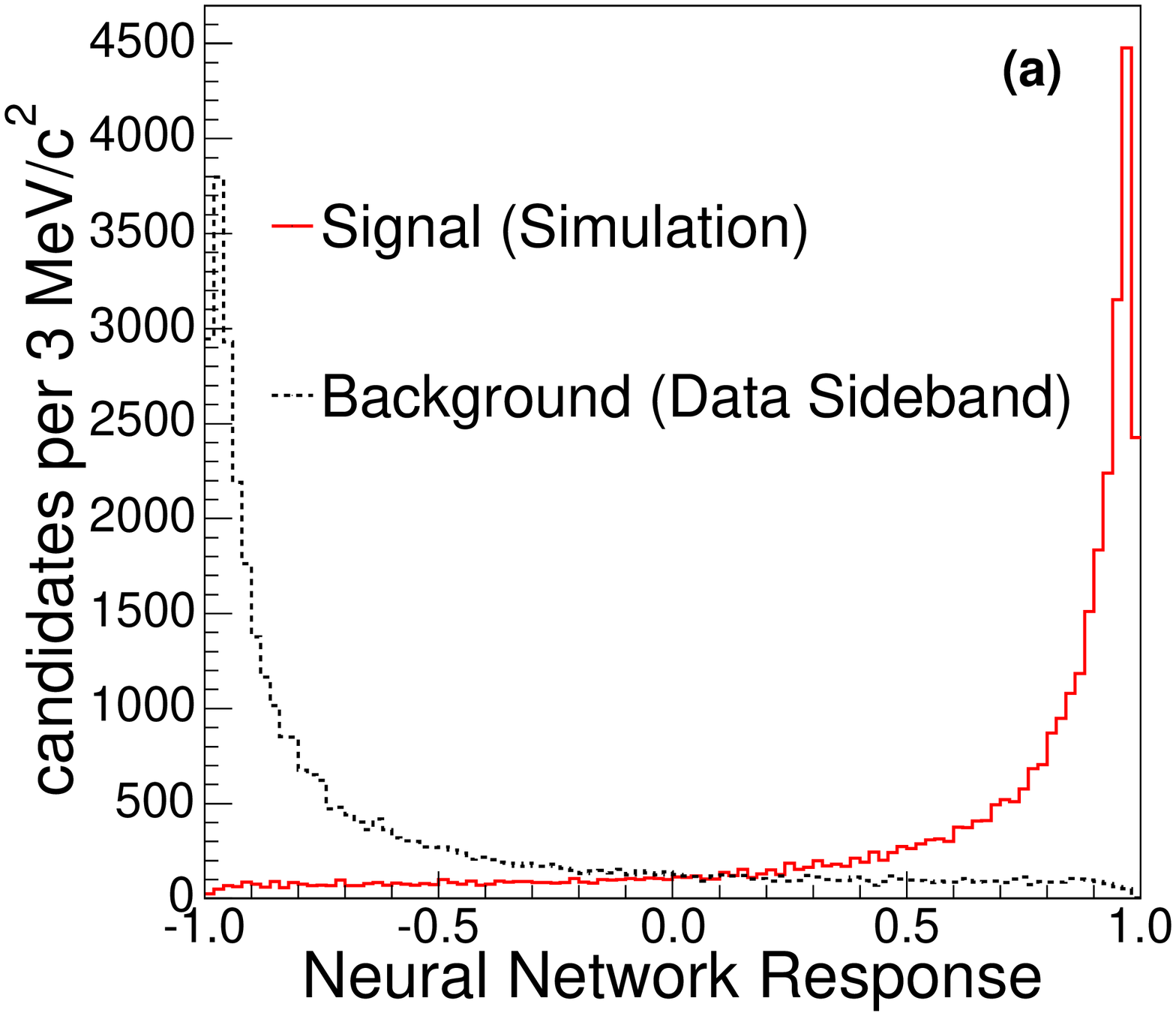}
    \includegraphics[angle=-90,width=8cm,clip=]{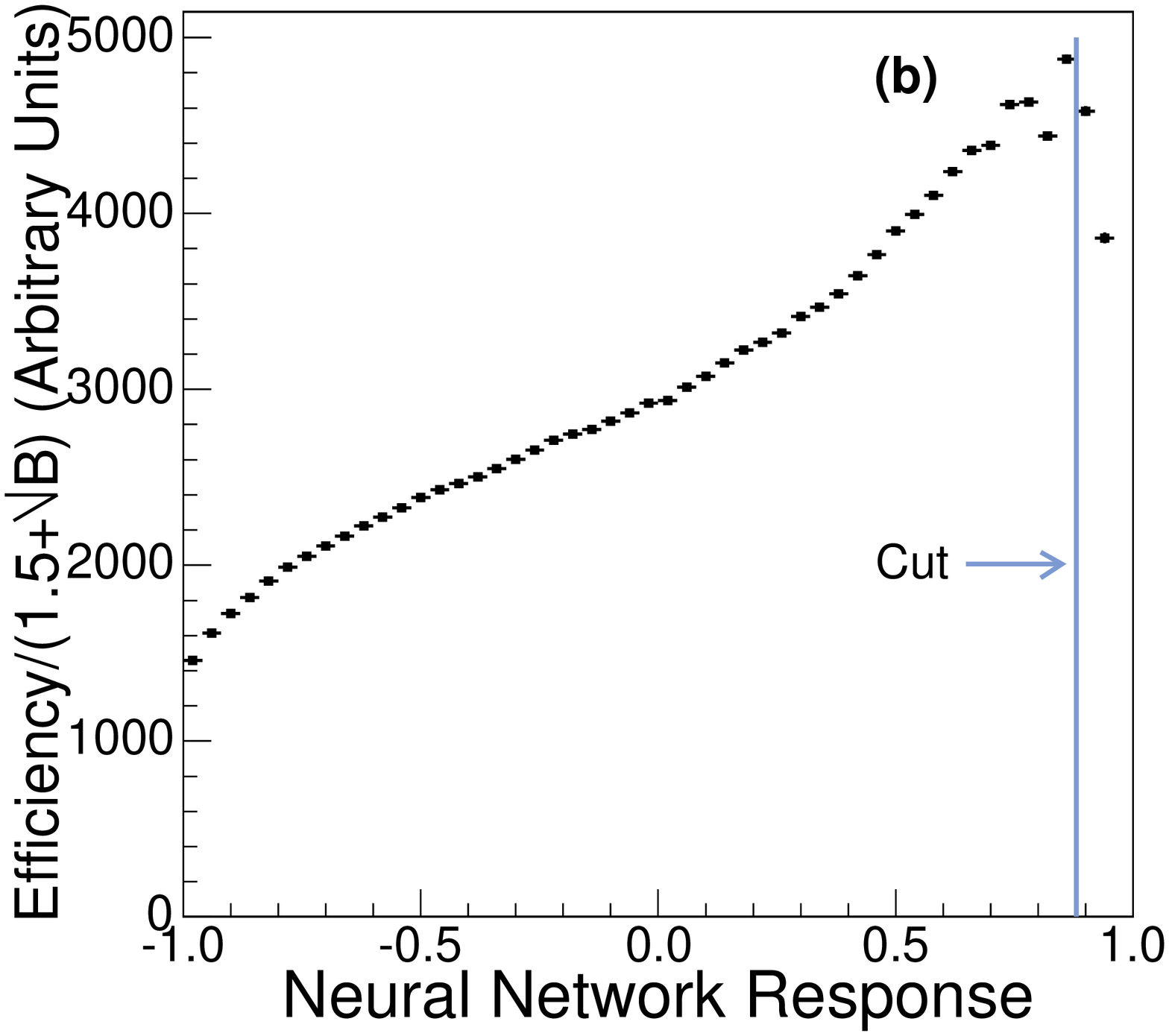}	
    \caption{(a) NN response where the solid line is signal simulation and the dashed one is sideband data. (b) Figure of merit $S/(1.5+\sqrt{B})$ as a function of NN response. The vertical line indicates the optimized cut in the NN response.\label{NNOuput}}
  \end{center}
\end{figure*}

We train the NN using simulated $B^0_s$ MC events as a signal sample. We use data from the upper sideband in the $J/\psi \, K^0_S$ candidate mass distribution, 
well separated from the signal region, as a background training sample. 
We use as inputs for the NN the quantities listed in Table~\ref{tabInputVariable}.  These input quantities are chosen 
as variables with good discriminating power which, alone or in combination, do not bias the mass spectrum. After the training, 
the NN achieves strong discrimination between signal and background as shown in Fig.~\ref{NNOuput}a.  

As in the $B^0_s\rightarrow J/\psi \, {K^*}^0$ analysis, we optimize the selection by maximizing $S/(1.5+\sqrt B)$. The signal $S$ is modeled using $B^0_s$ MC events in the reconstructed mass range 5.350 GeV/$c^2< M_B <$ 5.400 GeV/$c^2$. The background $B$ is modeled using $J/\psi \, K^0_S$ candidates in data populating the 
 mass range 5.430 GeV/$c^2< M_B <$ 5.480 GeV/$c^2$. The figure of merit suggests a cut value in the NN response of 0.88 as shown in Fig.~\ref{NNOuput}b.

\begin{figure*}[!p]
  \includegraphics[angle=-90,width=8cm,clip=]{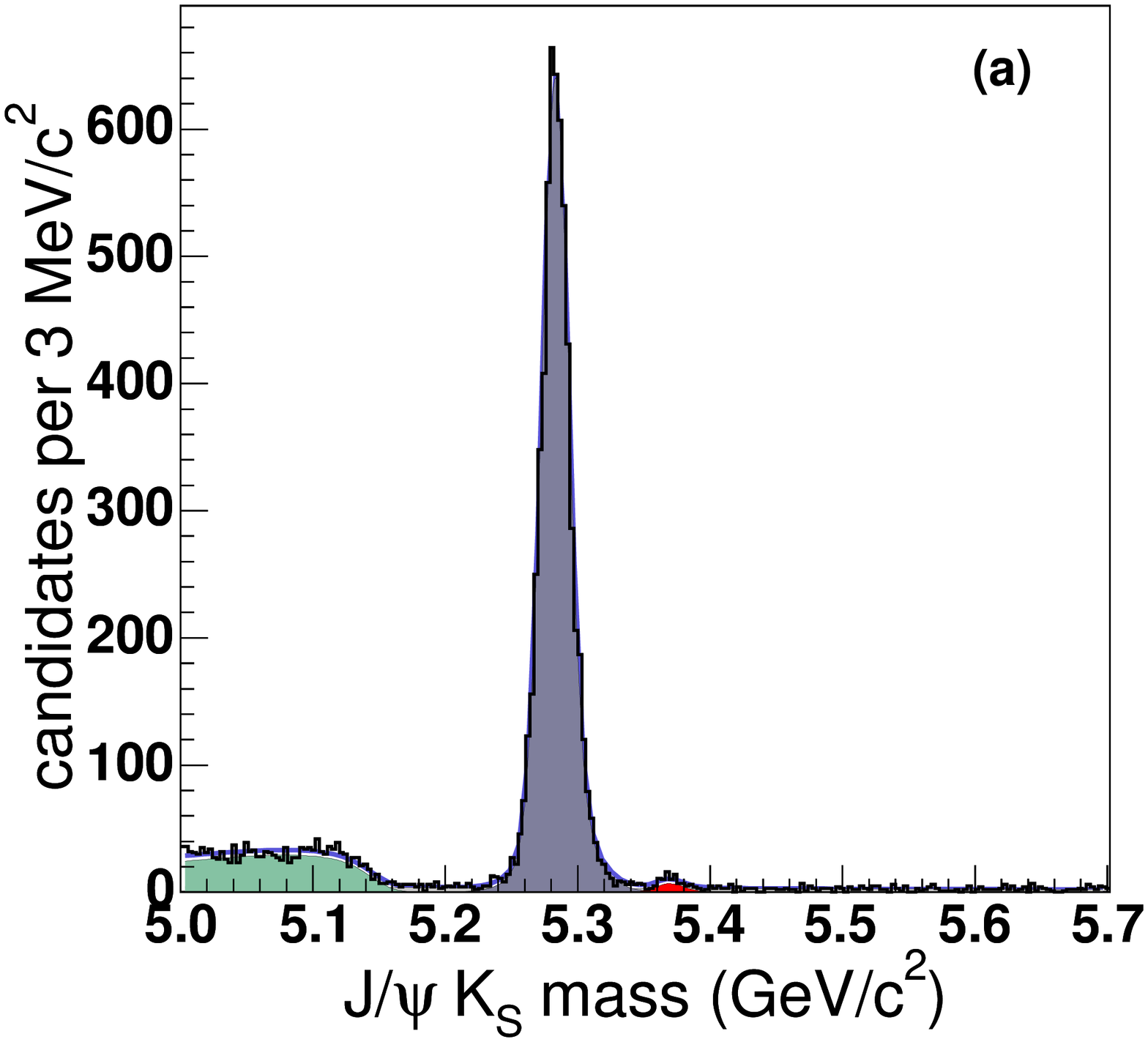}
   \includegraphics[angle=-90,width=8cm,clip=]{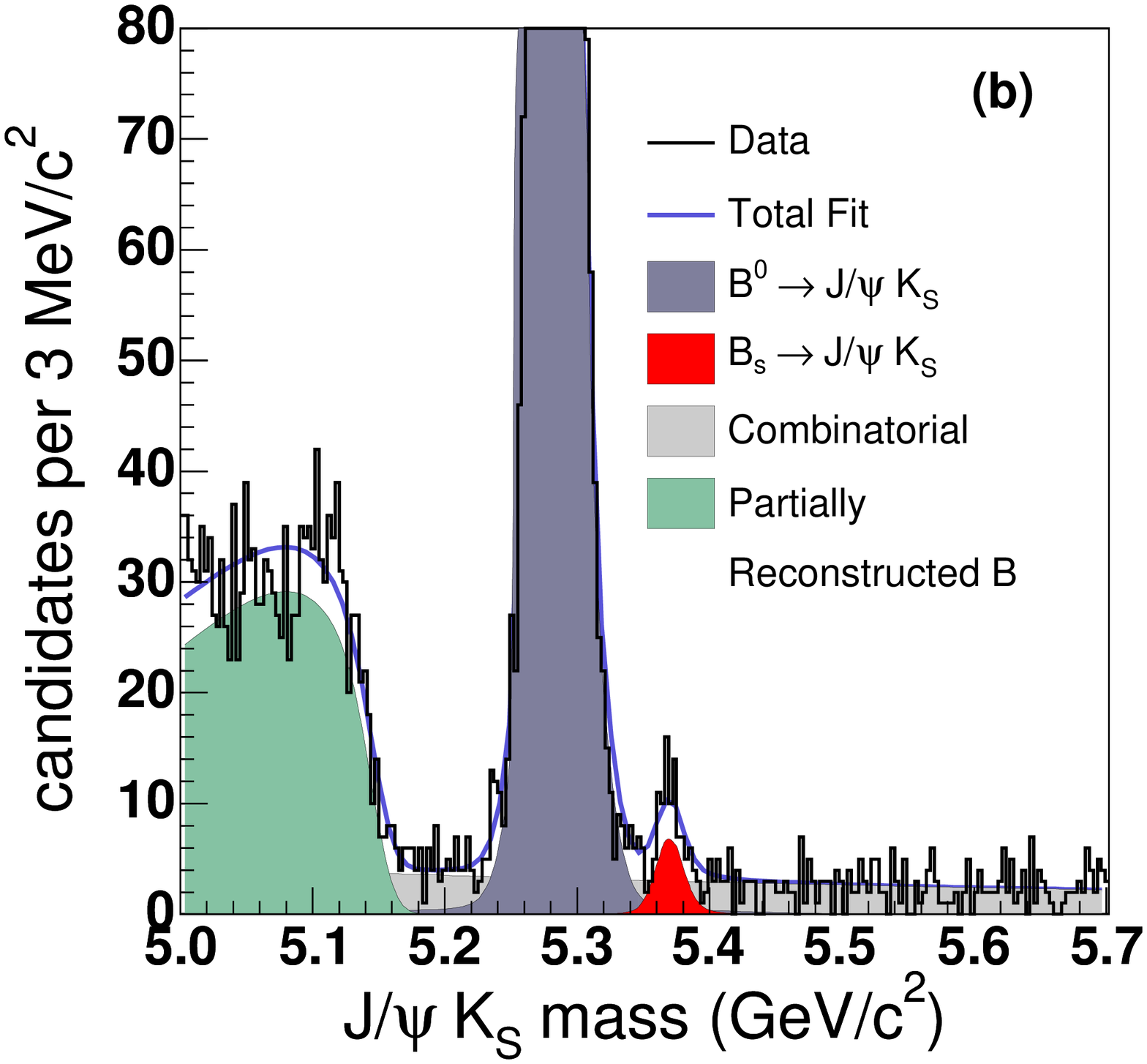}	
 \caption{\label{fig1}(a) Invariant mass distribution in data for $J/\psi \, K^0_S$ candidates and fit including the different contributions. (b) We enlarge the distribution in the signal region for more detail.}
\end{figure*}

The fitting technique is similar to the $B^0_s\rightarrow J/\psi \, {K^*}^0$ analysis. We obtain the yields of $B^0 \rightarrow J/\psi \, K^0_S$ and $B^0_s \rightarrow J/\psi \, K^0_S$ signals in a binned likelihood fit to the invariant mass distribution. We again model the $B^0$ and $B^0_s$ signal contributions with three Gaussian templates obtained from fitting $B^0\rightarrow J/\psi \, K^0_S$ MC and use the mass difference between $B^0_s$ and $B^0$ for the formation of the $B^0_s\rightarrow J/\psi \, K^0_S$ template. The two major sources of background in this analysis are combinatorial background and partially reconstructed $b$-hadron decays. We model these with the same functional forms used in the $B^0 \rightarrow J/\psi \, {K^*}^0$ analysis. However, we include only one ARGUS function because the contribution of partially reconstructed $B^0_s$ is negligible. An additional background in this analysis is ${\Lambda_b}^0 \rightarrow J/\psi \, \Lambda$ decays where the $p$ from the $\Lambda$ decay is assumed to be a $\pi$. In order to suppress the ${\Lambda_b}^0$ contribution, we apply a cut to the angular variable cos($\theta_{K^0_S,\pi_2}$), where $\theta_{K^0_S,\pi_2}$ is the angle between the $K^0_S$ candidate $p_T$ in the lab frame and the lower $p_T$ pion ($\pi_2$) in the $K^0_S$ center-of-mass frame. Cutting out events with cos($\theta_{K^0_S,\pi_2}$)$<$ -0.75 removes 99.8\% of the ${\Lambda_b}^0$ while retaining 86\% of the $B^0_{s}$. The residual  ${\Lambda_b}^0$ contamination is less than one event and is neglected. The invariant mass distribution for $J/\psi \, K^0_S$ and the fit result including the different contributions are shown in Fig.~\ref{fig1}.

We determine the yields of the $B^0 \rightarrow J/\psi \, K^0_S$ and $B^0_s \rightarrow J/\psi \, K^0_S$ signal to be 5954 $\pm$ 79 and 64 $\pm$ 14 respectively. 
As with the $B^0_s \rightarrow J/\psi \, {K^*}^0$ case, we determine the statistical significance of the
$B^0_s \rightarrow J/\psi \, K^0_S$  signal 
 by fitting the mass distribution without the $B^0_s$ contribution (background-only hypothesis), a difference of one degree of freedom between the two hypotheses. For likelihood ${\cal L}$ we interpret $-2\log{\cal L}$ as a $\chi^2$ and use the difference in that quantity to determine that the probability of background fluctuations producing a comparable or greater signal is 3.9$\times$10$^{-13}$ or 7.2$\sigma$.
 The value of $N(B^0_s \rightarrow J/\psi \, K^0_S)/N(B^0 \rightarrow J/\psi \, K^0_S)$ is 0.0108 $\pm$ 0.0019 (stat). 

The sources of 
systematic uncertainty are similar to the other analysis. In this case the absolute uncertainties for the ratio are 6$\times$10$^{-4}$ from the combinatorial background contribution, 6$\times$10$^{-4}$ from the combinatorial background modeling, 5$\times$10$^{-4}$ from the signal modeling and 1.3$\times$10$^{-5}$ from the mass difference between $B^0$ and $B^0_s$. The systematic uncertainties are summarized in Table~\ref{tabSysYields}. We sum the contributions in quadrature resulting in a total systematic uncertainty of $\pm$0.0010. The final value of $N(B^0_s \rightarrow J/\psi \, K^0_S)/N(B^0 \rightarrow J/\psi \, K^0_S)$ is 0.0108 $\pm$ 0.0019 (stat) $\pm$ 0.0010 (syst).

\section{Acceptance Calculation}
To determine the ratio of branching ratios ${\mathcal B}(B^0_s \rightarrow J/\psi \, K)/{\mathcal B}(B^0 \rightarrow J/\psi \, K)$, 
where $K$ represents $K^0_S$ or ${K^*}^0$, the relative 
acceptances of $B^0 \rightarrow J/\psi \, K^0_S$ to $B^0_s \rightarrow J/\psi \, K^0_S$ and 
$B^0 \rightarrow J/\psi \, {K^*}^0$ to $B^0_s \rightarrow J/\psi \, {K^*}^0$ need to be determined. We use MC samples to extract $A_{rel}$ as follows:

\begin{equation}
A_{rel}=\frac{N(B^0\rightarrow J/\psi \, K~{\rm pass})/N(B^0\rightarrow J/\psi \, K~{\rm gen})}{N(B^0_s\rightarrow J/\psi \, K~{\rm pass})/N(B^0_s\rightarrow J/\psi \, K~{\rm gen})},
\end{equation}

\noindent
where $N$(gen) is the number of MC generated signal events, $N$(pass) is the number of events passing all selection requirements, and $K$ represents $K^0_S$ or ${K^*}^0$. 

We determine the value for $A_{rel}$ to be 1.057 $\pm$ 0.010 for the ${K^*}^0$  channel 
and 1.012 $\pm$ 0.010 for the $K^0_S$ channel. We determine the statistical uncertainty on the acceptances for $B^0$ and $B^0_s$, assuming binomial statistics.  This MC statistical uncertainty is reported as a systematic 
uncertainty on $A_{rel}$.

The data sample utilized in this analysis was acquired using a number of variations on the 
$J/\psi \rightarrow \mu^+\mu^-$ trigger.  We have verified that the acceptance calculation is robust and consistent across
all kinematic variations of these triggers.
 
Several other effects contribute to the systematic uncertainty on $A_{rel}$. Uncertainty in $B^0_s$ and $B^0$ lifetimes introduce an uncertainty 
on the acceptance through the transverse decay length requirement.
For $B^0_s \rightarrow J/\psi \, K^0_S$ analysis, we generate different MC samples, varying the 
lifetimes by one standard deviation with respect to their measured values. We use the average measured value for $B^0$ and the evaluated $\tau_{B^0_{sH}}$ value for $B^0_s$~\cite{pdg}. The maximum deviation of $A_{rel}$ is 0.028, and we take this value as a systematic uncertainty.

For the $B^0_s \rightarrow J/\psi \, {K^*}^0$ analysis, the procedure to evaluate the systematic uncertainty is slightly different.  The  
$B^0_s \rightarrow J/\psi \, {K^*}^0$ decay  is an unknown admixture of 
$\mathit{CP}$-even and $\mathit{CP}$-odd states which have different lifetimes. 
The world-average currently gives $\Delta \Gamma_{B^0_s} / \Gamma_{B^0_s}$ = 0.092$_{-0.054}^{+0.051}$ for $\Gamma_{B^0_{s}}=\frac{1}{2}(\Gamma_{B^0_{sH}} +\Gamma_{B^0_{sL}})$~\cite{pdg}, where $\Gamma_{B^0_{sH}}$ and $\Gamma_{B^0_{sL}}$ are the widths 
of the heavy and light mass eigenstates respectively. If the $B^0_s$ were either all $B^0_{sH}$ or $B^0_{sL}$, the maximum lifetime change would be 5\%. To evaluate the effect on $A_{rel}$, we reweight the default $B^0_s \rightarrow J/\psi \, {K^*}^0$ lifetime distribution.  The reweighting is performed by normalizing the default lifetime distribution and comparing it to distributions with the lifetime increased or decreased by 5\%.  This leads to a maximum deviation on $A_{rel}$ of 0.046.

Another source of systematic uncertainty arises from the momentum spectra of the $B^0$ and $B^0_s$.   Since we normalize
our $B^0_s$ signal to the $B^0$ mode, we are sensitive only to mismodeling in the ratio of $p_T(B^0)$ versus $p_T(B^0_s)$, which should be quite small.
We compare the default $B^0_s$ and $B^0$ samples which use a next-to-leading order QCD calculation~\cite{nde} to the $p_T$ spectrum measured by CDF~\cite{cdf}. 
In the $B^0_s \rightarrow J/\psi \, {K^*}^0$ analysis, 
the value of $A_{rel}$ varies by 0.029 when 
using these alternative production spectra and we take this value as a systematic uncertainty.  Likewise, for the $B^0_s \rightarrow J/\psi \, K^0_S$ analysis, the change in $A_{rel}$ is 0.032.

Our relative acceptance is calculated assuming that the polarization in $B^0_s \rightarrow J/\psi K^*$ is identical to
the polarization in $B^0 \rightarrow J/\psi K^*$. Since we have no {\it a priori} knowledge of the actual polarization in 
the $B^0_s$ mode, we compute the systematic uncertainty by allowing all possible values for the polarization.  
We generated MC samples for $A_{0} = 1$, $A_{\parallel}=1$, and $A_{\perp}=1$.  The maximum variation from any of these polarizations leads to a systematic uncertainty on $A_{rel}$ of 0.261.   Since the angular distributions arising from polarization are clearly the dominant systematic
uncertainty, we have neglected the correlation between polarization and lifetime in assessing the uncertainties.  
 
Table~\ref{tabSysA} shows a summary of the systematic uncertainties on $A_{rel}$ for both measurements.  
Summing these contributions in quadrature, we find $A_{rel}$ = 1.057 $\pm$ 0.010 (stat) $\pm$ 0.267 {\rm (syst)} for the ${K^*}^0$    
 analysis and $A_{rel}$ = 1.012 $\pm$ 0.010 (stat) $\pm$ 0.042 {\rm (syst)} for the $K^0_S$ analysis.

\section{Results}
Using the values of $A_{rel}$ described above, we find

\begin{equation}
 \begin{split}
 \frac{f_s{\mathcal B}(B^0_s \rightarrow J/\psi \, {K^*}^0)}{f_d{\mathcal B}(B^0 \rightarrow J/\psi \, {K^*}^0)}&= 0.0168 \pm 0.0024 {\rm (stat)} \\
                    &\pm 0.0068 {\rm (syst)}
\end{split}
\end{equation}
and
\begin{equation}
 \begin{split}
 \frac{f_s{\mathcal B}(B^0_s \rightarrow J/\psi \, K^0_S)}{f_d{\mathcal B}(B^0 \rightarrow J/\psi \, K^0_S)}&= 0.0109 \pm 0.0019 {\rm (stat)} \\
&\pm 0.0011 {\rm (syst)}.
\end{split}
\end{equation}

%The ratio of branching ratios and the absolute branching ratios can be calculated using the results above and information available from other 
%measurements from CDF and other experiments. 
%Using measured values for $f_s/f_d$, we determine the ratio of branching ratios The ratio of branching ratios can be determined from these results using $f_s/f_d$. 
To determine the ratio of branching ratios, 
we combine these results with the 
most recent CDF measurement~\cite{fracCDF} of $f_s/(f_u+f_d)\times {\mathcal B}(D_s \rightarrow \phi \pi )$ and $f_u/f_d$ with the 
current world-average value~\cite{pdg} for ${\mathcal B}(D_s \rightarrow \phi \pi )$ to yield $f_s/f_d$ = 0.269 $\pm$ 0.033. 
We quote the systematic 
uncertainty coming from the $f_s/f_d$ uncertainty as ``frag''. The ratio of branching fractions to the reference $B^0$ decays are:

\begin{equation}
\begin{split}
 \frac{{\mathcal B}(B^0_s \rightarrow J/\psi \, {K^*}^0)}{{\mathcal B}(B^0 \rightarrow J/\psi \, {K^*}^0)}&= 0.062 \pm 0.009 {\rm (stat)} \\
     &\pm 0.025 {\rm (syst)} \pm 0.008 {\rm (frag)} 
\end{split}
\end{equation}
and
\begin{equation}
\begin{split}
 \frac{{\mathcal B}(B^0_s \rightarrow J/\psi \, K^0_S)}{{\mathcal B}(B^0 \rightarrow J/\psi \, K^0_S)}&= 0.041 \pm 0.007 {\rm (stat)} \\
 &\pm 0.004 {\rm (syst)} \pm 0.005 {\rm (frag)}. 
\end{split}
\end{equation}
The relative branching ratios observed for both modes are in good agreement with the expectation based upon the pure spectator model.

We use the world-average values for ${\mathcal B}(B^0 \rightarrow J/\psi \, {K^*}^0)$ and ${\mathcal B}(B^0 \rightarrow J/\psi \, K^0)$~\cite{pdg} for normalization 
to calculate the absolute branching fractions:

\begin{equation}
 \begin{split}
 {\mathcal B}(B^0_s \rightarrow J/\psi \, {K^*}^0)&= (8.3 \pm 1.2 {\rm (stat)}\pm 3.4 {\rm (syst)} \\
 &\pm 1.0 {\rm (frag)} \pm 0.4 {\rm (norm)})\times10^{-5} 
\end{split}
\end{equation}
and
\begin{equation}
 \begin{split}	
 {\mathcal B}(B^0_s \rightarrow J/\psi \, K^0)&= (3.5\pm 0.6 {\rm (stat)} \pm 0.4 {\rm (syst)} \\
&\pm 0.4 {\rm (frag)} \pm 0.1 {\rm (norm)})\times10^{-5}.
\end{split}
\end{equation}

In conclusion, we present the first observation and branching ratio measurement of the 
Cabibbo suppressed decays \BsJpsiKstar \ and \BsJpsiKshort.  With larger data samples and additional analysis, these modes 
can be used to further explore the properties of the $B^0_s$ system.

\section*{ACKNOWLEDGMENTS}

We thank the Fermilab staff and the technical staffs of the participating institutions for their vital contributions. 
This work was supported by the U.S. Department of Energy and National Science Foundation; the Italian Istituto 
Nazionale di Fisica Nucleare; the Ministry of Education, Culture, Sports, Science and Technology of Japan; 
the Natural Sciences and Engineering Research Council of Canada; the National Science Council of the Republic of China; 
the Swiss National Science Foundation; the A.P. Sloan Foundation; the Bundesministerium f\"ur Bildung und Forschung, 
Germany; the Korean World Class University Program, the National Research Foundation of Korea; the Science and Technology 
Facilities Council and the Royal Society, UK; the Institut National de Physique Nucleaire et Physique des Particules/CNRS; 
the Russian Foundation for Basic Research; the Ministerio de Ciencia e Innovaci\'{o}n, and Programa Consolider-Ingenio 2010, 
Spain; the Slovak R\&D Agency; and the Academy of Finland.

\vspace{5cm}

\begin{table*}[!p]
\begin{center}
\caption[]
{{Selection criteria for $B^0 \rightarrow J/\psi \, K$ candidates and $B^0_s \rightarrow J/\psi \, K$ candidates, where $K$ represents ${K^*}^0$ or $K^0_S$.\label{tabCutsKS}}}
\begin{tabular}{l c c}\hline\hline
 Variable (Units)                               & $B^0_s \rightarrow J/\psi \,{K^{*}}^0$  & $B^0_s \rightarrow J/\psi \,K^0_S$ \\\hline   
% $B^0$ Candidate mass (GeV/$c^2$)                      & $>$ 3                                   & $>$ 3\\
%                                                       & $<$ 6.7                                 & $<$ 6.7\\
 $B^0$/ $B^0_s$ candidate four-track fit $\chi^2$      & $<$ 50                                  & - \\
 $B^0$/ $B^0_s$ candidate four-track fit probability   &   -                                     & $>$ 10$^{-5}$ \\
 $B^0$/ $B^0_s$ candidate transverse momentum $p_T$ (GeV/$c$)   & $>$ 6                                   & $>$ 4 \\ 
 $B^0$/ $B^0_s$ candidate impact parameter ($\mu$m)             & $<$ 50                                  & -     \\
 $B^0$/ $B^0_s$ candidate transverse decay length significance $L_{xy}/\sigma$ &      -                   & $>$ 2 \\ 
 $J/\psi$ candidate mass (GeV/$c^2$)                   & $>$ 3.05                                & $>$ 2.8 \\
                                                       & $<$ 3.15                                & $<$ 3.3\\
 $J/\psi$ candidate 3D two-track fit $\chi^2$   & $<$ 30                                  & $<$ 30 \\ 
 $K$ candidate mass (GeV/$c^2$)                        & $>$ 0.55                                & $>$ 0.55\\
                                                       & $<$ 0.846                               & $<$ 0.846\\
 $K$ candidate 3D two-track fit $\chi^2$      & $<$ 30                                    & $<$ 20\\ 
 $K$ candidate transverse decay length $L_{xy}$ (cm) & -                                         & $>$ 0.5\\
 $\mu$ transverse momentum $p_T$ (GeV/$c$)             & $>$ 1.5&                                $>$ 1.5\\
 $\Delta \phi$ between the two muons (radians)         & $<$ 2.25                                & $<$ 2.25\\ 
 $\mu_1$ charge~$\times$~$\mu_2$ charge                & $= -1$                                    & $= -1$ \\
 $\Delta z$ in the beam line between the two $\mu$  (cm) & $<$ 5                                 & $<$ 5\\
 $\pi$ transverse momentum $p_T$ (GeV/$c$)             & -                                       & $>$ 0.5\\\hline\hline                
\end{tabular}

\end{center}
\end{table*}

\begin{table*}[!p]
\begin{center}
\caption[]
{{Systematic uncertainties for the ratio of yields.  All numbers in percent.\label{tabSysYields}}}
\begin{tabular}{ c c c} \hline\hline
%Source of                 & Relative Uncertainty for~~~           & Relative Uncertainty for \\ 
%Systematic Uncertainties  & $\frac{N(B^0_s \rightarrow J/\psi \, {K^*}^0)}{N(B^0 \rightarrow J/\psi \, {K^*}^0)}~(\%)$ & $\frac{N(B^0_s \rightarrow J/\psi \, K^0_S)}{N(B^0 \rightarrow J/\psi \, K^0_S)}~(\%)$\\ \hline 
Source  & $\delta \frac{N(B^0_s \rightarrow J/\psi \, {K^*}^0)}{N(B^0 \rightarrow J/\psi \, {K^*}^0)}~(\%)$ &~ $\delta \frac{N(B^0_s \rightarrow J/\psi \, K^0_S)}{N(B^0 \rightarrow J/\psi \, K^0_S)}~(\%)$\\ \hline 
Signal Modeling                          & 4.4                       & 4.6  \\ 
Mass difference between $B^0$ and $B^0_s$  & 0.1                       & 0.1  \\ 
Combinatorial background modeling        & 1.3                       & 5.6  \\ 
Combinatorial background contribution    & 31.4                      & 5.6  \\ 
$B^0_s\rightarrow J/\psi \, \phi$ contribution & 1.3                 & -  \\ \hline
Total                                     & 31.8                     & 9.2\\\hline\hline
\end{tabular}

\end{center}
\end{table*}

\begin{table*}[!p]
\begin{center}
\caption[]
{{Variables used as input in the NN training.\label{tabInputVariable}}}
\begin{tabular}{l}\hline\hline
~~~~~~~~~~~~~~~~~~Input variables in the NN \\\hline
$B^0$/$B^0_s$ candidate transverse momentum \\
$B^0$/$B^0_s$ candidate four-track decay point fit \\
$B^0$/$B^0_s$ candidate proper decay length \\ 
$B^0$/$B^0_s$ candidate impact parameter\\
$J/\psi$ candidate transverse momentum \\
$J/\psi$ candidate mass\\
$J/\psi$ candidate proper decay length\\ 
$J/\psi$ candidate impact parameter\\
$K^0_S$ candidate transverse momentum\\
$K^0_S$ candidate mass\\
$K^0_S$ candidate proper decay length\\ 
$K^0_S$ candidate impact parameter\\
$\pi$ transverse momentum\\
$\pi$ impact parameter \\
$\mu$ transverse momentum \\
$\mu$ impact parameter \\
$\mu$ cosine of the helicity angle in $J/\psi$ rest frame \\\hline\hline
\end{tabular}

\end{center}
\end{table*}

\begin{table*}[!p]
\begin{center}
\caption[]
{{Systematic uncertainties for the relative acceptances. All numbers listed in percent.  \label{tabSysA}}}
\begin{tabular}{c c c} \hline\hline
%Source of                 & Relative Uncertainty for $A_{rel}$~~~  & Relative Uncertainty for $A_{rel}$\\ 
%Systematic Uncertainties   & in $B^0_s \rightarrow J/\psi \, {K^*}^0~(\%) $ & in $B^0_s \rightarrow J/\psi \, K^0_S~(\%)$\\ \hline

Source   & $\delta A_{rel}$($B^0_s \rightarrow J/\psi \, {K^*}^0~(\%)$) 
& ~~~~~$\delta A_{rel}$($B^0_s \rightarrow J/\psi \, K^0_S~(\%)$)\\ \hline
Lifetime for $B^0$ and $B^0_s$      & 4.4                    & 2.8  \\	
$B$ hadron $p_T$ spectrum           & 2.7                     & 3.2  \\
Polarization                      & 24.7                    & -  \\ \hline
Total                             & 25.3                    &  4.2\\\hline\hline
\end{tabular}

\end{center}
\end{table*}


\begin{thebibliography}{0}
\bibitem{eta} The CDF reference frame uses cylindrical coordinates, where $\theta$ and $\phi$ are the polar and azimuthal angles with respect to the proton beam. Pseudorapidity $\eta$ is defined as -ln(tan($\theta$/2)).  Transverse momentum $p_T$ is the charged particle momentum measured in the plane perpendicular to the beamline.
\end{thebibliography}
\end{document}